# Meta-Optics with Lithium Niobate

**Anna Fedotova[1,2], Luca Carletti[3], Attilio Zilli[4], Frank Setzpfandt[2,5], Isabelle Staude[1,2], Andrea Toma[6], Marco Finazzi[4], Costantino De Angelis[3], Thomas Pertsch[2,5], Dragomir N. Neshev[7†], and Michele Celebrano[4†]**

[1] Institute of Solid State Physics, Friedrich Schiller University – Jena, Germany

[2] Institute of Applied Physics, Abbe Center of Photonics, Friedrich Schiller University – Jena, Germany

[3] University of Brescia, Department of Information Engineering and INO-CNR – Brescia, Italy

[4] Politecnico di Milano, Department of Physics – Milano, Italy

[5] Fraunhofer Institute for Applied Optics and Precision Engineering, Jena, Germany

[6] Istituto Italiano di Tecnologia – Genova, Italy

[7] ARC Centre of Excellence for Transformative Meta-Optical Systems (TMOS), Research School of Physics, Australian National University, Canberra ACT 2601, Australia

†Corresponding author: dragomir.neshev@anu.edu.au, michele.celebrano@polimi.it

# Abstract

The rapid development of metasurfaces — 2D ensembles of engineered nanostructures — is presently fostering a steady drive towards the miniaturization of many optical functionalities and devices to a subwavelength size. The material platforms for optical metasurfaces are rapidly expanding and for the past few years, we are seeing a surge in establishing meta-optical elements from high-index, highly transparent materials with strong nonlinear and electro-optic properties. Crystalline lithium niobate (LN), a prime material of choice in integrated photonics, has shown great promise for future meta-optical components, thanks to its large electro-optical coefficient, second-order nonlinear response and broad transparency window ranging from the visible to the mid-infrared. Recent advances in nanofabrication technology have indeed marked a new milestone in the miniaturization of LN platforms, hence enabling the first demonstrations of LN-based metasurfaces. These seminal works set the first steppingstone towards the realization of ultra-flat monolithic nonlinear light sources with emission ranging from the visible to the infrared, efficient sources of correlated photon pairs, as well as electro-optical devices. Here, we review these recent advances, discussing potential perspectives for applications in light conversion and modulation shaping as well as quantum optics, with a critical eye on the potential setbacks and limitations of this emerging field.

## 1. Introduction

In the last decade, the field of meta-optics received a major boost, driven by the urge for the realization of ultra-compact photonic devices to shape and manipulate optical electromagnetic radiation [1] [2] [3] [4] [5] [6] [7] [8]. The prefix 'meta', which means 'after' in the Greek language but is currently employed mostly with the meaning of 'beyond' in science and technology, thoroughly entails the main aim of this research field: *i.e.,* pushing the capabilities and the integration of optical devices beyond that of conventional optics (*e.g.*, lenses, mirrors, filters, nonlinear crystals). The key enabling platforms at the core of this field are optical metasurfaces, sub-micrometer thick devices based on engineered ensembles of interacting nanostructures (*i.e.* meta-atoms), which promise to complement and enrich the set of tools already available in optics and photonics [2] [4] [5] [6] [7] [8]. This intriguing perspective is underpinned by the possibility to tailor the optical properties of metasurfaces through the combined design of the individual meta-atoms and of their ensemble. In particular, optical metasurfaces can be engineered to simultaneously feature a variety of fundamental optical functionalities, such as polarization control and wavefront shaping (*e.g.* light focusing, beam shaping) in a single ultra-thin device [3] [8] [9]. This is very appealing in photonic applications that require a high level of integration, such as in ultra-compact laser cavities (*e.g.*, vertical cavity surface-emitting laser) and integrated photonics [10]. Given their extremely compact design, metasurfaces can be also stacked in multiple serial elements to attain even more flexibility or add further functionalities [7]. The broad range of possibilities of this fascinating research topic attracted the focus of the photonic community in the last couple of decades. Thus far, this revolution has mainly concerned the field of passive optics [1] [2] [3] [4] [5] [6] [7] [8], which entails refractive elements capable of manipulating the light wavefront by addressing the spatial and temporal phase of the propagating beam.

The field of active meta-optics is devoted to the development of meta-devices capable of processing light beyond the refractive and diffractive regimes. Despite its rapid development triggered by the fascinating promises it holds [11] [12] [13], active meta-optics remains sizably unexplored to date. The main aim of this field is to provide additional functionalities to metasurfaces, such as nonlinear light conversion, dynamic light modulation and steering as well as generation of entangled photon pairs. Importantly, all the above processes are based on either photon-electron, photon-photon or photon-phonon interactions, which ultimately are mediated by matter via resonant or non-resonant electronic transitions. The fundamental challenge to overcome in this research area is the dramatic drop of light-matter interaction caused by the negligible material volume and, hence, the sub-wavelength light propagation. Indeed, optical metasurfaces are conceived to interact with and manipulate propagating light. As such, they constitute open photonic systems that weakly interact with the electromagnetic radiation, which makes their exploitation as active media extremely challenging. The common materials of choice for the technological development of active and nonlinear photonics are semiconductors, such as Silicon (Si), Gallium Arsenide (GaAs), or Gallium Phosphide (GaP), and ionic crystals, such as lithium niobate (LN), beta barium borate (BBO), or potassium dihydrogen phosphate (KDP) [14]. In particular, the extensive use of the latter is motivated, along with sizable nonlinear optical coefficients, by a broad transparency window that extends down to the visible wavelengths. This allows them to sustain the high peak powers required to obtain sizeable light-matter interaction. In bulk active optics, the employment of long propagation lengths along with crystal poling allows maximizing the phase matching condition between the interacting fields, allowing to further increase light-matter interaction [14].

Although the inapplicability of phase matching in meta-optics undoubtedly decreases the overall efficiency of the processes, the possibility of circumventing it lifts the constraints on momentum and energy conservation, therefore potentially enabling broadband nonlinear light conversion, as recently demonstrated in 2D materials [15]. Yet, to obtain efficient electro-optical modulation and nonlinear optical processes in active meta-optics, alternative approaches have to be devised to enhance light-matter interaction at lengths confined much below the operating wavelength. Plasmonic-based (*i.e.* metallic) metasurfaces were the first platforms proposed to tackle these challenges in meta-optics, because of their ability to enhance light-matter interaction and attain extremely intense localized fields at sub-wavelength scales [2] [11] [12] [16]. Still, the utmost drawback of these platforms – when operating at optical wavelengths – are the large dissipative losses in metals, which considerably limit the employed optical powers and, hence, the overall efficiency of the underlying processes.

The advances in state-of-the-art nanofabrication of CMOS-compatible platforms steered recent development of semiconductor-based all-dielectric platforms for meta-optics [17] [18] [19]. To date, these metasurfaces have been successfully applied to nonlinear light conversion and demonstrated their potential in light-by-light modulation and active light steering. In particular, group IV semiconductors have been used for third-harmonic generation (THG), [20] [21] [22] whereas III−V semiconducting alloys have been employed for second-harmonic generation (SHG) [23] [24] [25] [26]. Yet, because of their relatively narrow band gaps with absorption onsets in the near-infrared (NIR) wavelength range, most of these platforms limit the development of low-loss nonlinear meta-optics, at least in a crucial region of the electromagnetic spectrum: the visible range. The meta-optics components therefore need to incorporate highly transparent, high-index, functional materials, which have driven a quest for the development of new material platforms, such as nanostructured ferroelectrics.

LN is among the most widespread materials employed in integrated photonics and nonlinear optics [27] [28], thanks to its extremely wide transparency window (350 – 4500 nm) along with its unique nonlinear- and electro-optical properties. Yet, the realization of LN-based meta-optics has only recently been theoretically [29] [30] [29] and experimentally [31] [32] [33] [34] [35] [36] [37]investigated. Despite the extended application in integrated optics [27] [38], the fabrication of monolithic platforms with sub-micron features is extremely challenging still to date. Nevertheless, it is already clear from the first experimental demonstrations that LN metasurfaces with reproducible nanometer-size features hold great promise for the deployment of nonlinear meta-optics down to the visible range.

Here, we will review this novel class of platforms and their application to active meta-optics, discussing the immediate implication to the field and the sought perspective in the medium-long term. In Section 2 we will discuss the general LN properties, while in Section 3 we will address the approaches nowadays available for the nanofabrication of LN nanostructures and metasurfaces, analyzing future directions in material engineering and nanofabrication technologies. In Section 4 we will focus on the recent advances in the realization of LN metasurfaces for nonlinear and electro-optical light management, describing the state of the art in light conversion, photon pair generation and electro-optical modulation capabilities. Finally, in section 5, we will describe the potential future directions and perspective application of these platforms in nonlinear optics, quantum optics, and electro-optics.

## 2. Properties of LN

LN is a dielectric uniaxial crystal, transparent over a large wavelength range spanning from the ultra-violet (UV) at 350 nm all the way to the mid-infrared (MIR) at 4500 nm. Being uniaxial, LN is birefringent, thus enabling advanced polarization control/manipulation. The birefringence of the material stems from the polar nature of its crystalline lattice, which additionally provides access to many peculiar functionalities like ferroelectricity, pyroelectricity and piezoelectricity, *i.e.*, the possibility of modifying the material properties via external stimuli such as electric fields, temperature and pressure variations, respectively.

However, the most prominent feature of LN for its application to active optics is the relatively high second-order nonlinearity, enabled by the broken inversion symmetry of the crystalline lattice.

In the technologically relevant telecom wavelength range (*i.e.,* around 1550 nm), LN features one of the higher nonlinearities among the ionic crystals. The largest element of the nonlinear tensorial coefficient *d* (by definition half of the nonlinear susceptibility tensor $\chi^{(2)}$) is $d_{33} = 21\ pm/V$ [39](see Table 1), which becomes even larger at shorter wavelengths (e.g. $27\ pm/V$ at 1064 nm [40]). This allows for efficient nonlinear parametric three-wave mixing, e.g., effects like SHG, sum- and difference-frequency generation (*i.e.,* SFG and DFG) as well as spontaneous parametric down-conversion (SPDC). The broken inversion symmetry in LN, along with large nonlinearity, enables a strong electro-optic effect, namely the Pockels effect. This effect, where an applied external field imparts a variation of the refractive index of the material, already displayed a significant technological impact in optics, integrated photonics, and laser technologies, allowing the realization of variable polarizer and dynamic light modulation for the implementation of phase modulators.

| Material | Refractive index ($n$) at $\lambda$ = 1550 nm) | Transparency window | Larger nonlinear coefficients/susceptibilities |
|---|---|---|---|
| Non-centrosymmetric materials | | | |
| Lithium Niobate (LiNbO$_3$) | 2.2 | 350 nm - 4.5 µm | $d_{33} \sim 21 \frac{pm}{V}$ [39][4] <br> ($d_{33} \sim 27 \frac{pm}{V}$ [40])[5] |
| (Aluminum) Gallium Arsenide ((Al)GaAs) | 3.5 | (700)[1] 900 nm – 18 µm | $d_{36} \sim 170 \frac{pm}{V}$ [41][1][5] |
| Gallium Phosphide (GaP) | 3 | 450 nm - 14 µm | $d_{36} \sim 70 \frac{pm}{V}$ [41][5] |
| Gallium Nitride (GaN) | 2.3 | 370 nm – 7 µm | $d_{33} \sim 10 \frac{pm}{V}$ [42][5] |
| Barium Titanate (BaTiO$_3$) | 2.3 | 360 nm – 9 µm | $d_{15} \sim 17 \frac{pm}{V}$ [6] [43] [44] |
| Halide Perovskites | 1.9 | >400 nm[2] | $d^{\text{eff}} \sim 1 \div 14 \frac{pm}{V}$ [45][3] |
| Centrosymmetric materials | | | |
| Germanium (Ge) | 4.2 | 1.8 µm – 20 µm | $\chi^{(3)} \sim 5.6 \times 10^{-19} \frac{m^2}{V^2}$ [14] |
| Silicon (Si) | 3.5 | 1000 nm – 8 µm | $\chi^{(3)} \sim 2.8 \times 10^{-18} \frac{m^2}{V^2}$ [14] |

**Table 1.** Linear and nonlinear optical properties of LiNbO3 compared to other widely used dielectric materials. <u>Notes</u>: (1) For aluminum concentration of about 20%. (2) The bandgap varies widely but, for most perovskites, it is above 700 nm. (3) Nonlinear coefficients strongly depend on the perovskite constituents. Values reported at (4) 633 nm, (5) 1060 nm, (6) 1550 nm.

These properties make LN very interesting for optical applications in general and for nano-optics in particular, where modulation and frequency conversion are recurring topics. Conversely, a significant challenge to the employment of LN in nano-optics is posed by the relatively low refractive index, which spans from 2.2 to 2.4 going from the near-infrared to visible spectral range. Yet, the aforementioned functionalities/properties clearly outweigh this last drawback in a large set of applications.

In Table 1, we compare the optical properties of LN with different other materials used for nonlinear nanooptics. Whereas semiconductor materials have a higher refractive index and 2$^{nd}$-order nonlinearity, most of them have limited application in the visible-UV spectral range due to the typically high absorption.

Furthermore, in particular the often-used III-V semiconductors feature a crystal structure that provides only nonlinearities coupling all three polarization components, which makes it challenging to exploit the high magnitude of the nonlinearity with co-propagating beams. Finally, LN provides unique means for fast dynamic modulation of the material properties by the electro-optic effect, which enables to tune the refractive index by an externally applied voltage with frequencies well-above the GHz [46].

## 3. Novel nanofabrication approaches for lithium niobate: new possibilities for a well-established material

LN as a crystalline material poses demanding challenges for nanofabrication, especially when the sought-after structures need to be fabricated in thin films, which thus far inhibited the deployment of LN in nanophotonics. Here we will describe the recent developments in LN nanofabrication that allows to address them and discuss new strategies to improve the nanofabrication and to enhance LN-based platforms.

### 3.1. State of the art and perspectives in lithium niobate nanofabrication

Very generally, we can divide nanofabrication methods into two major categories: top-down, when we deterministically remove parts of the material of interest, or bottom-up, when we let chemicals assemble into micro- and/or nanostructures. LN is chemically very stable [47], so top-down methods are difficult to apply. The major milestone for LN nanophotonics was the fabrication of thin films with the smart-cut (crystal ion-slicing or ion-cut) technology, which finally reached the market in the middle of 2010s [48] [49]. This brought LN back to the attention of various research groups focused on nanophotonics. The main idea of smart cut relies on the implantation of $He^+/H^+$ ions in a well-localized thin layer at a specific depth within the LN substrate. In this layer, the crystal structure is modified, which creates an interface where the wafer can be eventually split into two pieces. The depth of the implantation defines the thickness of the future thin film, which typically is in the range of 400 nm to 1000 nm. The implanted LN wafer is then bonded on a $SiO_2$ film placed on a Si or a LN substrate. After thermal annealing, the original LN wafer can be split along the "cut" plane. With this method it is possible to fabricate thin films with different orientation of the crystal axes, where the extraordinary crystalline c-axis can be oriented normal to the thin film (*z*-cut) or parallel to it (*x*-cut or *y*-cut).

For microscale devices based on thin films, such as microdisks, waveguides, modulators, etc. there are many fabrication techniques available such as femtosecond-laser micromachining, dicing, proton-exchange, ion implantation, and metal diffusion [33] [50] [51] [52] [53] [54] [55] [56]. Nonetheless, they still lack the spatial resolution needed to produce LN meta-atoms small enough to support resonances at optical frequencies. Within this context, focused-ion-beam milling (FIB) and electron-beam-lithography (EBL) based processes have already shown very convincing results for the fabrication of metasurfaces [31] [32] [33] [32] [57] [58] [31] [59].

FIB milling uses ions (typically $Ga^+$), accelerated to energies of around 30 keV, to bombard the surface of a material and sputter it as sketched in Figure 1a). This approach is relatively straightforward, does not require a mask, yields high resolution (10-15 nm) and produces nanostructures with steep side walls (~86°, see Figure 1c)) [33] [60]. However, FIB is based on point-wise ablation, which is time consuming and expensive, thus not suitable for patterning areas exceeding 100 x 100 $\mu m^2$. Other problems are surface roughness, which can deplete the resonator performances, and even more crucial, $Ga^+$ contamination, which disturbs the LN crystal lattice and introduces defects. These problems can be reduced by optimizing the FIB operating parameters and by introducing a sacrificial/protective layer that also has the function of preventing charge accumulation and can be eventually removed in a post-processing step [33].

EBL-based techniques rely on multi-step protocols, similar to those implemented for silicon photonics or electronic applications (see sketch in Figure 1d). A typical process starts with applying a hard mask and a photoresist on the wafer to be structured. Then the electron beam shapes the photoresist, and in the next step a lift-off process or reactive ion etching (RIE) can be used to structure the hard mask. Finally, the mask

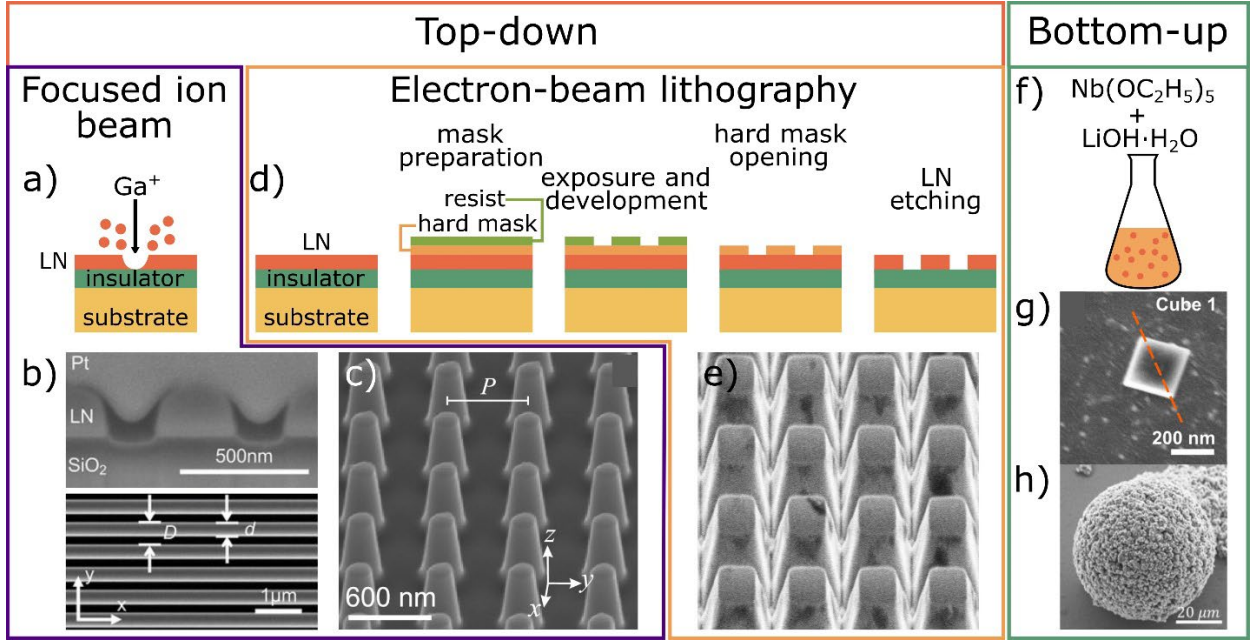

**Figure 1**. **Overview of fabrication methods and examples of fabricated structures.** (a-e) Top-down methods based on (a-c) focused ion beam (FIB) and (d-e) electron-beam lithography (EBL). (f-h) Bottom-up methods. (a) Sketch of the core idea behind the FIB fabrication. Examples of structures obtained by FIB milling: (b) 1D metasurface with nanograting, top – cross-section with FIB, bottom – SEM image of the nanograting [31]; (c) SEM of a 2D metasurface [33]. (d) Sketch with the main steps of EBL-based fabrication. (e) Example of a fabricated structure: SEM of a 2D metasurface [32]. (f) Sketch including the essential chemicals for solvothermal method. (g) SEM pictures of a nanocube fabricated by solvothermal method [64] and (h) crystalline microspheres fabricated by hydrothermal method (unpublished, courtesy of the authors of Ref.[66])

pattern can be transferred to the LN layer via ion-beam etching (IBE), inductively-coupled plasma RIE (ICP-RIE), or RIE [32] [61]. EBL can process large sample areas, which makes it more suitable for wafer-scale processing than FIB. Typical minimal structure size would be around 20 nm and the typical side-wall angle for metasurfaces varies between 70-80°, but both these parameters as well as surface roughness strongly depend on the exact fabrication recipe. To boost the efficiency of nanoresonators further improvements of fabrication techniques are necessary. Indeed, while in silicon metasurfaces the highest reported quality factor (Q-factor) reaches $10^5$ [62], only very recently LN metasurfaces have attained values up to $10^3$ [63].

Alternatively to the described techniques based on thin films and deterministic structuring, single nanoparticles sustaining resonances can also be fabricated with bottom-up approaches [64] [65] [66] [67].

This can be achieved with laser ablation [65], but also several possibilities to produce (crystalline) LN particles with chemical synthesis were demonstrated: molten salt syntheses, sol–gel methods, and solid-state approaches [68] [69] [70] [71] [72]. Unfortunately, in these techniques it is hard to control the exact size and shape of the grown nanoparticles, which is of crucial importance for designing nanostructures resonating at specific wavelengths. Precise positioning of nanoparticles is also very difficult. Hence, for applications relying on deterministic location of nanoresonators, FIB- and EBL-based approaches still represent the most promising and reliable techniques.

## 3.2. Enhancing lithium niobate

LN is routinely used as a material for integrated optics, where several strategies to further enhance its capabilities were already introduced [38] [73] [74] [75], some of which can also profitably be used in nanophotonics.

### 3.2.1. Doping

Doping crystals such as LN with other materials can be used to change their properties [76]. For example, doping with MnO induces variations in energy levels in LN, pushing the absorption even further to the UV or to the MIR [77] or changing the refractive index [78]. However, the refractive-index variations are typically of the order of $10^{-3}$ and will not sizably change the characteristics of LN in nanophotonic applications, which are typically based on a large refractive-index difference.

Transition-metal ions like Fe, Cu, Mn are used as dopants to increase the photorefractive effect, i.e. the change in the refractive index due to irradiation with light [79] [80] [81]. This could be used to realize dynamically tunable nanophotonic elements. On the other hand, LN doping with metal ions like Mg, Zn, In, Sc, Hf, Zr, Sn strongly reduces photorefraction enabling radiation-stable optical elements by reducing the optical damage induced by lasers [82]. Fabricating metasurfaces from films with these dopants will increase their damage resistance especially when used with CW lasers.

Rare-earth dopants like Er, Yb, Tm, on the other hand, can act as fluorescence emitters when embedded in LN, extending the material functionality [74] [75]. Many works exploited Er and Yb for lasing in LN microdisks and waveguides [83] [84] [85] [86] [87]. Metasurfaces from LN have the potential to decrease the lasing threshold, control and direct the emission. Furthermore, such ions can act as single-photon emitters and employed to embed quantum light sources into LN.

### 3.2.2. Periodic poling

LN is a ferroelectric crystal, which possesses a spontaneous polarization, i.e., its polarization is nonzero even in the absence of an external electric field. This is a consequence of the lack of inversion symmetry of the lattice, which leads to the separation of positive and negative charges. When a strong electric field is applied in the direction of the crystalline *c*-axis, it is possible to displace the Li and Nb ions and thus to permanently invert the spontaneous polarization in a process called electric-field poling. This also leads to a sign change in the nonlinear susceptibility $\chi^{(2)}$. Based on this effect, second-order nonlinear effects in LN can be controlled, e.g., by applying this procedure to some parts of a LN thin film to create (a)periodic poling patterns, which results in a structured $\chi^{(2)}$-nonlinearity. This is especially interesting for x-cut or y-cut films as the *c*-axis falls into the plane of the film, thus facilitating the onset of nonlinear processes seeded by propagating beam impinging at normal incident with respect to the device.

Periodic poling of LN thin films was already employed for waveguides, microrings, etc. to support quasi-phase-matching for second-harmonic generation [56] [88] [89]. Moreover, recent advances in the poling algorithms demonstrated submicron poling periods [90]. State-of-the-art works achieve poling periods as small as 750 nm for *x*-cut [91] [92] and 300 nm for *z*-cut LN thin film [93]. Using electric field poling, one can engineer the nonlinearity and add a degree of freedom for manipulation of the fields generated in frequency-conversion processes also for nanophotonics and metasurfaces.

## 4. Lithium niobate metasurfaces

### 4.1. Nonlinear optics with lithium niobate metasurfaces

The technological advances in nonlinear optics in the last half century enabled an entire realm of novel photonic devices and groundbreaking applications, which already brought a major contribution to science and technology (laser physics, material science, quantum optics, etc.), as well as in our society (theragnostic, nanomedicine, biology, etc.). Yet, the field of nonlinear meta-optics is still in its infancy. Indeed, although the impact of nonlinear metasurfaces in the field is clear [12], to date the nonlinear conversion efficiency attainable is far from bulk materials due to the negligible material volume in these photonic systems. As already mentioned, although semiconductor-based (*e.g.*, Si, Ge, GaP, GaN, GaAs) metasurfaces are unambiguously establishing as key platforms in all-dielectric nonlinear meta-optics [13] thanks to the recent advances in nanofabrication technologies, the operation of many of these meta-devices

The unique properties of LN described in Section 2, which already granted a broad diffusion and application in linear and nonlinear integrated photonics [27] [38], make it an ideal solution also for active meta-optics, especially for nonlinear optics and electro-optics applications. The most intense nonlinear optical effects as well as the electro-optical effects in non-centrosymmetric materials are governed by the nonlinear susceptibility tensor $\chi^{(2)}$ (or the simplified nonlinear tensor $d = \frac{1}{2}\chi^{(2)}$, as a result of the Kleinman symmetry condition), which mixes the impinging electric field components at frequencies $\omega_2$ and $\omega_1$ generating a nonlinear polarization vector at $\omega_3$:

$$\begin{pmatrix} P_x^{(2)}(\omega_3) \\ P_y^{(2)}(\omega_3) \\ P_z^{(2)}(\omega_3) \end{pmatrix} = 4\varepsilon_o \begin{pmatrix} d_{11} & d_{12} & d_{13} & d_{14} & d_{15} & d_{16} \\ d_{21} & d_{22} & d_{23} & d_{24} & d_{25} & d_{26} \\ d_{31} & d_{32} & d_{33} & d_{34} & d_{35} & d_{36} \end{pmatrix} \begin{pmatrix} E_x(\omega_1)E_x(\omega_2) \\ E_y(\omega_1)E_y(\omega_2) \\ E_z(\omega_1)E_z(\omega_2) \\ E_y(\omega_1)E_z(\omega_2) + E_z(\omega_1)E_y(\omega_2) \\ E_x(\omega_1)E_z(\omega_2) + E_z(\omega_1)E_x(\omega_2) \\ E_y(\omega_1)E_x(\omega_2) + E_x(\omega_1)E_y(\omega_2) \end{pmatrix} \quad (4.1)$$

where $\omega_3 = \omega_1 + \omega_2$, while the quadratic nonlinear tensor in the near infrared is [40]:

$$d = \begin{pmatrix} 0 & 0 & 0 & 0 & -4.3 & -2.1 \\ -2.1 & 2.1 & 0 & -4.3 & 0 & 0 \\ -4.3 & -4.3 & -27 & 0 & 0 & 0 \end{pmatrix} \frac{pm}{V}. \quad (4.2).$$

Unlike other quadratic nonlinear materials (e.g. III-V semiconductors), which are commonly used to realize optical metasurfaces, LN is a trigonal crystal with $3m = C_{3v}$ symmetry, which is unique given it has only three non-zero components in its nonlinear tensor (see Eq. 3.2), where $d_{33}$ is the dominant quadratic nonlinear contributions in the nonlinear tensor [40] [94]. The possibility to easily cut the material along a preferential direction, hence defining a specific crystal axis orientation provides a further degree of freedom to engineer the nonlinear optical processes in LN-based platforms. As previously mentioned, phase matching conditions, which are exploited in bulk and guided optics to the maximize the efficiency of nonlinear optical processes, do not apply to meta-optics. This introduces the key challenge of efficiency enhancement in nonlinear meta-optics, but concurrently grants higher flexibility – *i.e.,* broadband nonlinear conversion [15]. An effective approach to overcome the lack of phase matching in these confined systems is to engineer the Mie resonances of the individual meta-atoms and lattice resonances stemming from their periodic arrangement. These key strategies, thus far hindered by the limited LN nanofabrication capabilities, are to date becoming accessible thanks to the recent developments described in Section 3. This allows to envisage the application of LN-based metasurfaces for the enhancement and modulation of the nonlinear light generated by both stimulated (see Subsection 4.1.1) and spontaneous (see Subsection 4.1.2) second-order nonlinear optical processes, with important implications in the field of ultra-compact devices for light conversion and quantum optics, respectively. Finally, the aforementioned strong electro-optic coefficient of the material, which already made LN a key player for integrated electro-optics in telecom applications, can be also exploited to enable nonlinear light modulation and steering by external driving fields (see Section 4.2).

The last years witnessed a major drive in the development of LN-based nonlinear meta-optics, supported by a significant number of designs and scientific publications on the subject. In the following we will revise this emerging field, focusing on the key advantages of the technology and the perspectives that it can bring about and critically addressing some fundamental drawbacks that need to be confronted before arriving at a full deployment of this technology.

### 4.1.1. Stimulated nonlinear processes

The nonlinear wave mixing in nanostructured materials can be described as interactions between the different resonant modes of the structure, excited at the different wavelengths. These are often addressed as the quasi-normal modes of the structure [95]. However, in contrast to the larger-scale microstructures, in nanostructured materials, such as metasurfaces, the modes contain all three electric field components (see Eq. 3.1), which are often of similar magnitude. The nonlinear polarizations generated by the mixing between these field components via the nonlinear susceptibility tensor $\chi^{(n)}$ of the material are the primary sources for the nonlinear optical signal. Below we mainly focus on the second-order nonlinear interactions, which are specific to the use of quadratic nonlinear materials such as LN.

Second harmonic generation (SHG) is usually the most prominent nonlinear process, which stems from the mixing of the two impinging fields mediated by $\chi^{(2)}$ tensor. In general, the overall efficiency of SHG is governed by the cross-coupling coefficient $\kappa$. Within nanoresonators, one has a complex field distribution, and $\kappa$ is a function of the spatial overlap integral between the electric field of the optical resonant mode ($E^\omega$) at the impinging frequency $\omega$ and that ($E^{2\omega}$) at the generated frequency $2\omega$ computed over the volume $V$ of the nanostructure

$$\kappa \propto \left| \sum_{ijk} \chi^{(2)}_{ijk} \int_V E_i^{2\omega*}(\mathbf{r}) E_j^\omega(\mathbf{r}) E_k^\omega(\mathbf{r}) \, d\mathbf{r} \right|^2 \qquad (4.1)$$

where the indices $i,j,k$ run over the Cartesian components $x,y,z$. In bulk and guided optics, $\kappa$ can be largely simplified, leading to the well-known phase-matching condition. Conversely, in nano-optics, $\kappa$ depends on the mode fraction that is contained within $V$. This, along with the joint presence of all the electric field components, makes the optimization of $\kappa$ one of the major tasks in nonlinear nano-optics. Effectively, $\kappa$ represents a mode matching integral, which at the nanoscale replaces the phase-matching conditions in bulk nonlinear crystals [96] [97]. As the fields inside the nanostructures are enhanced proportionally to the quality factor of the resonances, the SHG efficiency can be expressed as $\eta^{SHG} \propto \kappa Q_{FW}^2 Q_{SH}$, where $Q_{FW}$ and $Q_{SH}$ are the quality factors at the fundamental wave and the second harmonic wave respectively. We note that the conversion efficiency is quadratically dependent on the quality factor of the fundamental wave resonance, hence the resonances matching the frequency of the pump field are the most critical for harmonic generation. Finally, since it is strongly influenced by the material and nanostructure symmetry along with the local density of states in the nanostructure, SHG emission is often characterized by highly nontrivial polarization and spatial properties. This was widely explored for individual LN nanostructures, including nanowires and waveguides [98] [99], nanocubes [64], nanospheres [65] and nanodisks [100].

Based on the discussion above, in ultra-thin metasurfaces maximizing the quality factors of the resonances and optimizing the mode overlaps are the keys for improving the harmonic generation efficiency. In particular, SHG is strongly enhanced by the excitation of resonances at the spectral position of the interacting waves [13]. However, we note that special care needs to be taken to ensure that the spectral bandwidth of the resonances equals or is larger than the bandwidth of the excitation pulses, in order to couple all of the incident energy into the resonant mode. Importantly, in metasurfaces there are additional constraints imposed by the symmetry of the underlying arrangement of the meta-atoms, which may alter the mode structure in the near field and spatially filter the individual meta-atom emission in the *k*-space, causing the formation of distinct second-harmonic diffraction orders and shaping the directionality of the emission [25] [26].

The first exploration of LN metasurfaces for enhancing SHG emission incorporated plasmonic resonant modes enabled by gold rings around small LN nanopillars [101], as shown in Figure 2a. The LN nanopillars were etched in a monolithic substrate of an *x*-cut LN crystal. The use of the *x*-cut material allowed to access the strongest nonlinear tensor component, namely $d_{33}$. SHG enhancement of nearly 20 times with respect to the unstructured substrate was observed, showing the potential of strong SHG enhancement by nanostructuring the LN material.

The development of thin-film LN by ion slicing for integrated nano-optics has sparked new opportunities for integration of LN with meta-optics. For instance, gradient silicon metasurfaces have been used to achieve phase matching-free SHG in integrated waveguides [10]. However, the big opportunity was in exciting resonant modes in the LN material itself without the need for further hybrid integration. By nanostructuring an *x*-cut LN thin film with the appropriate dimensions, electric and magnetic-type dipolar resonances were excited in a metasurface at telecommunication wavelength, see Figure 2b [32]. The largest SHG efficiency was observed for the resonance dominated by the electric contributions because its specific field distribution enables the most efficient usage of the largest nonlinear tensor element, $d_{33}$.

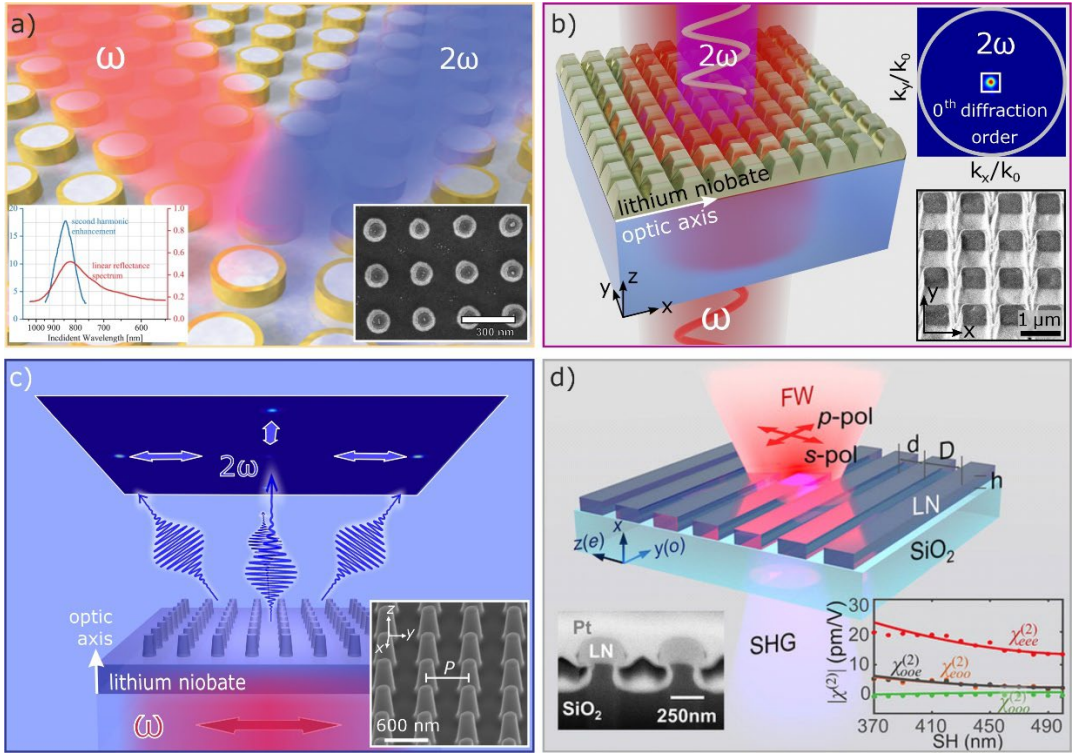

**Figure 2**: **Nonlinear metasurfaces based on LN for SHG enhancement.** (a) Plasmonic nanoring metasurface integrated on *x*-cut LN substrate to enhance SHG [101]. (b) Mie-resonant LN metasurface by nano structuring a thin *x*-cut LN film [32] and (c) *z*-cut [33] LN substrates. w and 2w denote the angular frequency of pump and SHG emission, respectively. The bottom-right insets display scanning electron micrographs of the metasurfaces. The top-right inset in a) and the dark blue frame in c) show back-focal plane images of the SHG emission in the direction space. (d) A LN-based nonlinear photonic grating to enhance and manipulate SHG in free-space optics [34].

A fundamental advantage of nanostructured metasurfaces is their ability to generate strong longitudinal fields, which is not possible in bulk crystals. As such, in optical metasurfaces it is possible to access the nonlinear tensor components that otherwise cannot be directly excited. For example, *z*-cut LN can be used to fabricate resonant metasurfaces with highly enhanced harmonic generation [33]. Figure 2d shows such a metasurface fabricated in a monolithic LN substrate and operating in the visible spectral range [33]. The SHG process in this metasurface was enhanced by a magnetic-type Mie-resonance and resulted in the emission of blue light at ~415 nm with no absorption. This metasurface exhibited a SHG conversion coefficient $\eta^{SHG} = P_{pk}^{SHG}/(P_{pk}^{exc})^2$ of up to $10^{-10}$ [W$^{-1}$], where $P_{pk}$ is the pulse peak power. While the *x*-cut metasurface designs showed strong zeroth order emission (see top-right inset in panel b), the *z*-cut metasurface emits predominantly in the first diffraction orders (panel c), whose relative power can be modulated via pump polarization as the emission is dominated by two diffraction orders oriented along it. Similar SHG at visible wavelengths has also been found in 1D Mie-resonant nanogratings based on *x*-cut LN thin film, as shown in Figure 2d [32]. The 1D structures show extreme polarization anisotropy and enhancement at different visible wavelengths, achieved by changing the width of the LN grating stripes. However, the intrinsic need for the Mie-resonant optical modes to couple to free space propagating light along with the moderate refractive index of LN result in relatively low-quality factors (Q~10) of the resonances. Such lower Q-factors are in stark contrast to LN integrated optical components, such as microdisk resonators [102] and photonic crystals [27] [103]. Therefore, the SHG enhancement in Mie-resonant LN metasurfaces is limited.

To further enhance the SHG conversion efficiency, it is possible to utilize *high quality factor (high-Q) modes* based on the concept of bound state in the continuum (BIC) [104]. This possibility has been theoretically investigated suggesting significant enhancement of the SHG process, while controlling the coupling to free-space radiation through symmetry breaking and excitation of quasi-BIC modes [30] [105] [106]. In particular, the concept of etchless LN metasurfaces has been introduced, which avoids the complex etching process of LN and circumvents the roughness of the side walls of etched LN metasurfaces [107]. This concept has been recently tested experimentally in *x*-cut etchless LN metasurfaces, demonstrating experimental quality factors of Q=455 and more than 50 times SHG enhancement in comparison to the bare LN film [108]. As the BIC modes consist of transversely propagating waves, they are strongly affected by the transverse size and edges of the metasurfaces.

| Sample | $\lambda_{FH}$ [nm] | $\lambda_{SHG}$ [nm] | $Q_{FH}$ | $I^{pk}_{FH}$ [GW/cm²] | $\eta$ ($P_{SHG}/P_{FW}$) | $\gamma$ ($P^{pk}_{SHG}/P^{pk\,2}_{FW}$) [W⁻¹] | $\xi=\eta/I^{pk}_{FH}$ [cm²/GW] |
|---|---|---|---|---|---|---|---|
| Experiments | | | | | | | |
| 2D MS [32] | 1550 | 775 | ~100 | 4.5 | 9.00E-07 | 2.43E-10 | 2.00E-07 |
| 2D MS [33] | 820 | 410 | ~10 | 0.5 | 2.53E-08 | 2.86E-11 | 5.05E-08 |
| 1D MS [34] | 820 | 410 | ~50 | 2.05 | 1.98E-06 | 1.39E-09 | 9.64E-07 |
| 1D DBR MS [63] | 1548 | 774 | ~2000 | 0.002 | 4.09E-08 | 3.02E-09 | 2.13E-05 |
| Simulations | | | | | | | |
| 2D MS [29] | 820 | 410 | ~20 | 1 | 5.00E-05 | - | 5.00E-05 |
| 1D BIC MS [30] | 690 | 345 | ~10000 | 1.33 | 8.13E-05 | - | 6.11E-05 |
| 2D stack BIC MS [105] | 1334 | 667 | ~1000 | 5.3 | 1.30E-04 | - | 2.45E-05 |
| 2D BIC MS [106] | 801 | 400.5 | ~80000 | 3.3E-06 | 4.90E-03 | - | 1.48E+03 |
| 2D MS on Au [109] | 800 | 400 | ~40 | 3.4 | 5.00E-04 | - | 1.47E-04 |

**Table 2.** SHG performances of LiNbO₃ metasurfaces. <u>Notes</u>: (1) 1D: one-dimensional; 2D: two-dimensional; MS: metasurface; DBR: distributed Bragg reflector; BIC: bound state in the continuum. (2) $\lambda_{FW}$: wavelength of the fundamental (pump) wave (FW). (3) $\lambda_{SH}$: wavelength of the second-harmonic (SH) emission. (4) $Q_{FW}$: quality factor at the FW. (5) $I^{pk}_{FW}$: pulse peak intensity of the FW field calculated as average input power ($P_{FW}$) divided by the spot size on the metasurface.

However, recent experiments have shown that the implementation of a heterostructure cavity can mitigate the transverse leakage of energy and further boost the Q-factors of the metasurfaces (to over 5000), thereby dramatically enhancing SHG processes, though at the expense of a reduced bandwidth [63]. In Table 2 we compare the experimental and simulative key works on SHG in LN metasurfaces. All the studies have been carried out in the last three years, hence demonstrating the strong drive towards the development LN metasurfaces due to its indisputable advantages.

### 4.1.2. Spontaneous nonlinear processes

Many quantum applications like quantum cryptography, quantum computing, quantum information require photon pairs, quantum states of light containing exactly two correlated photons. To generate such pairs, mainly the second-order nonlinear process spontaneous parametric down-conversion (SPDC) is used [110]. In SPDC, one pump photon spontaneously decays into two daughter photons, signal and idler, controlled again by energy conservation and the field overlap in the source. However, while phase matching allows to use propagating modes interacting with large volumes of the nonlinear material to reach meaningful generation rates, this becomes a fundamental challenge in nano- and meta-optics due to the intrinsically low amount of material involved as discussed before.

For SPDC, nanostructures offer a broader angular and spectral emission [111] [112] (see Figure 3a) of the generated photon pairs as well as a smaller footprint [38]. First experiments in a 300 nm $x$-cut LN thin film showed a photon-pair rate of 0.3 Hz/mW for pump wavelength 685 nm [113]. Interestingly, the measured spectral width of the SPDC photons was more than 500 nm, as shown in Figure 3b, and was limited only by the experimental setup.

To leverage the full potential of nanooptics for photon-pair generation, the next step are nanostructured metasurfaces as sources for these quantum states. Several works already suggested to use metasurfaces for generation, manipulation and detection of quantum states [19] [114] [115], some of them concentrating on III-V semiconductors (GaAs, AlGaAs) [116] [117] [118] and silicon [119].

These works showed, that in quantum optics metasurfaces offer similar benefits compared to thin films or bulk crystals as in classical nonlinear optics: they show resonant field enhancement, enable wavefront manipulation, etc. Resonances play a crucial role here as they provide enhanced density of states and, thus, potentially enhance SPDC. So far researchers concentrate on designing metasurfaces with resonances for signal and idler wavelengths.

Before we discuss metasurfaces with nanostructured LN, we highlight etchless hybrid approaches where non-structured LN thin films are combined with other resonant structures on top of the LN film, which can provide localized resonances, field enhancement, and shaping of the propagating waves [108] [116] [120]. Such a metasurface with $SiO_2$ grating on top (see Figure 3c) had a resonance at 1570.5 nm with bandwidth 3.5 nm (at normal incidence). It demonstrated degenerate entangled photons with telecom wavelengths generated via SPDC with photon-pair generation rate 1.1 μHz/(mW*μm^3) [108]. After varying pump wavelength and, thus, moving away from the grating's resonance, the SPDC rate went down as shown in Figure 3d. Interestingly, the SPDC spectrum was theoretically estimated to be narrow, 3 nm, which probably is a result of the enhancement of the density of states for the resonance wavelength only. Another work suggested using silver nanogratings on LN film on silver substrate to generate localized surface and gap plasmons [120]. They should provide strong electric field enhancement within the LN layer. The bottom silver layer acted as a perfect reflector and thus, SPDC photons acquired strong directionality and were detected from the grating side. Two grating resonances at 1479 nm and 1254 nm yielded signal and idler photons into these wavelengths. The calculated non-degenerate SPDC rate reached 0.46 mHz/(mW*µm$^3$) at 679 nm. However, metals suffer from losses which decreases the conversion efficiency.

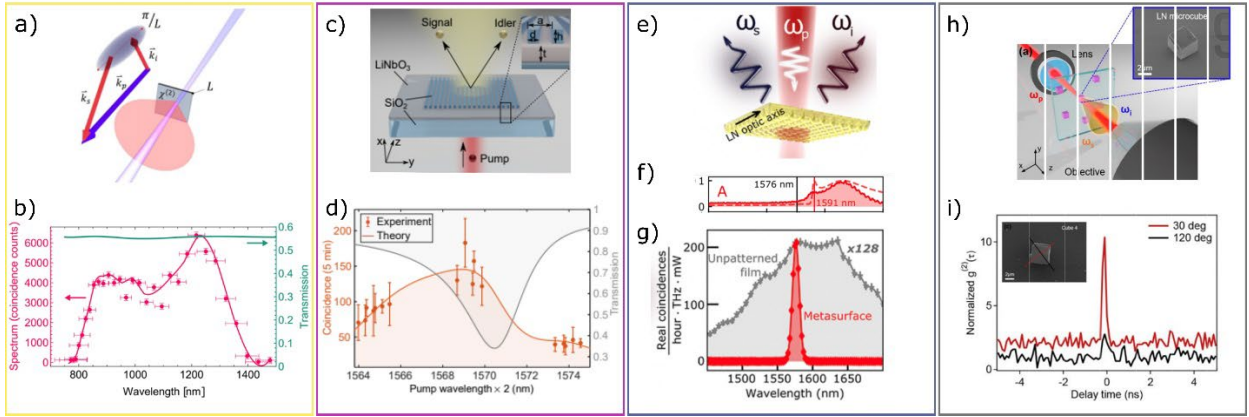

**Figure 3.** SPDC in LN – state of the art. a) Illustration of relaxed phase-matching condition and broadband SPDC generation from subwavelength nonlinear films [111]. b) Measured SPDC spectrum of photon pairs generated from 300 nm x-cut LN thin film when pumping at 515 nm (pink dots) with pink line as a fit for detection efficiency. Calculated transmission of Fabry-Perot etalon formed by LN film (green line) [113]. c) Scheme of SPDC generation in transmission from a non-local metasurface with ~300 nm x-cut LN thin film and $SiO_2$ grating. d) Coincidence histograms of SPDC for different pump wavelengths from 780 nm to 790 nm (orange dots) and numerically calculated SPDC via quantum–classical correspondence with sum-frequency generation (orange line). Grey line – transmittance spectrum of the metasurface showing resonance at 1570.5 nm [108]. e) Scheme of SPDC generation in reflection from a LN metasurface with LN thickness 680 nm (fabricated from x-cut LN thin film). f) Measured and calculated reflectance spectra of a LN metasurface with sharp resonance at 1591 nm. g) Measured SPDC spectrum from the metasurface with resonance at 1591 nm (red) and from unpatterned LN thin film (grey) when pumping at 788 nm [121]. h) Scheme of SPDC generation in transmission from a single LN nanocube with size 4 μm. i) Coincidence histogram of SPDC for two incident pump polarizations: red – along LN optic axis, black – perpendicular to it. Inset – SEM picture of the nanocube, the red line indicates the LN optic axis [122].

Moving to LN metasurfaces, in Ref. [121] the SPDC from several different resonant metasurfaces was studied. The maximum obtained generation rate was 0.5 mHz/(mW*µm$^3$) for the most efficient metasurface when pumping at 788 nm (Figure 3e). The most intriguing feature, however, were the generated SPDC spectra, whose width could be controlled by the spectral difference between twice of the pump wavelength and the resonance wavelength (see Figures 3f-g). This work thus demonstrates the potential to control the SPDC spectra by tuning the resonance positions.

| Platform | Thickness (µm) | Power (mW) | SPDC rate[1] Hz/(mW·µm) | SPDC rate[2] mHz/(mW·µm$^3$) | SPDC spectral width |
|---|---|---|---|---|---|
| LiNbO$_3$ film [111] | ~6 | 220 | 1.06 | 13.5 | 500 nm |
| LiNbO$_3$ film [113] | 0.3 | 9 | 1.04 | – | 500 nm @515 nm pump |
| LiNbO$_3$ metasurfaces [121] | 0.68 | 70 | 0.11 | 6.1 | 10÷40 nm[3] |
| LiNbO$_3$ slab + SiO$_2$ grating [108] | 0.304 | 85 | 0.07 | 0.009 | 3 nm (theory) |
| LiNbO$_3$ nanocube [122] | 3.6 | 50 | 0.007 | 0.57 | – |
| LiNbO$_3$ slab + silver grating (simulations) [120] | 0.306 | 5[4] | 915[5] | 1.17[5] | 1 nm[5] |

**Table 3.** Summary of measured and numerically calculated SPDC properties. <u>Notes</u>: (1) Rate normalized to pump power and LiNbO$_3$ thickness. (2) Rate normalized to pump power and LiNbO$_3$ volume. (1,2) are detected rates, losses in collection/detection are not accounted for. (3) The spectral width depends on the detuning of the pump from the resonance. (4) Calculated from intensity 1 W/cm$^2$ for the metasurface with size 1 mm$^2$. (5) SPDC rate is normalized to the bandwidth of 1 nm.

Another approach for photon-pair generation is using single resonant structures like LN particles. In Ref. [122], a single resonant 4 µm LN nanocube generated photon pairs with rate 3.3 mHz/(mW*µm$^3$) when pumped with ~780 nm (see Figures 3h-i). Here we can observe the importance of resonances: this particular cube shows much better efficiency also because it has resonances at the pump and SPDC wavelengths. However, the nanocubes were fabricated using bottom-up solvothermal synthesis, and it is hard to control their optic axis direction and sizes.

Table 3 summarizes all current works on SPDC from LN nanostructures and metasurfaces. Although we normalized the published photon-pair generation rates on pump power and volume of LN in the fourth column to access how the structures perform, the values are still not fully comparable as not all the works mention detection efficiency and propagation losses. Also, we did not take SPDC spectral width in corresponding experiments into account.

## 4.2. Electro-optics with lithium niobate metasurfaces

The electro-optic (EO) effect is the most well-known nonlinear optical phenomenon exploited in LN and the reason for the large success of this material for integrated photonic circuits. This second-order nonlinear process is based on the variation of the refractive index in the presence of a static (or low-frequency) electric field. The amplitude of the refractive index modulation is described, using the contracted notation, as:

$$\begin{bmatrix} \Delta\left(\frac{1}{n^2}\right)_1 \\ \Delta\left(\frac{1}{n^2}\right)_2 \\ \Delta\left(\frac{1}{n^2}\right)_3 \\ \Delta\left(\frac{1}{n^2}\right)_4 \\ \Delta\left(\frac{1}{n^2}\right)_5 \\ \Delta\left(\frac{1}{n^2}\right)_6 \end{bmatrix} = \begin{bmatrix} 0 & 0 & 0 & 0 & r_{42} & r_{22} \\ -r_{22} & r_{22} & 0 & r_{42} & 0 & 0 \\ r_{13} & r_{23} & r_{33} & 0 & 0 & 0 \end{bmatrix} \begin{bmatrix} E_x \\ E_y \\ E_z \end{bmatrix},$$

where the electro-optic coefficients, at a wavelength of 500 nm, are $r_{33} \sim 31$ pm/V, $r_{13} \sim 8$ pm/V, $r_{22} \sim 4$ pm/V, $r_{42} \sim 28$ pm/V, and $E_x$, $E_y$, and $E_z$ are the amplitudes of the vectorial components of the static electric field [123] [124]. As it can be noted, the refractive index modulation linearly depends on the field amplitude. The effect is maximized when the field is oriented parallel to the crystal axis, in this case:

$$\begin{cases} \Delta n_e = \frac{1}{2} n_e^3 r_{33} E_z \\ \Delta n_o = \frac{1}{2} n_o^3 r_{13} E_z \end{cases}$$

Harnessing this phenomenon has enabled the development of waveguide-based integrated electro-optic modulators that are critical for high-speed telecommunication networks [125] [126]. Electro-optic modulation is potentially also attractive for devising dynamic new metastructures compatible with free-space optical components operating at GHz modulation rates [127] and it is currently attracting interest [35] [36] [37]. Indeed, a modulation of the refractive index can potentially spectrally shift a resonant mode and, consequently, modify the optical response of a metasurface. However, the resulting modulation of the refractive index is limited by breakdown, which in LN corresponds to 65 kV/mm but reduces to 3 kV/mm in air [128]. Thus, the achievable refractive index modulation amplitudes in the visible-near-infrared spectrum are on the order of $10^{-4}$. In the context of electro-optic waveguide-based modulators, this issue is addressed by using long propagation lengths [125] [126]. On the other hand, in resonant-based structures as metasurfaces, one needs to use optical modes with high Q-factor. So far, the EO effect was investigated theoretically and experimentally in either one-dimensional [35] [36] or two-dimensional [37] [129] metasurfaces for the realization of dynamic ultra-flat devices. All these realizations are based on thin LN films where the modulating electric field produced by two planar electrodes is aligned along the metasurface plane and uses the $r_{33}$ component, which is the strongest. In Ref. [35], the EO effect is used to vary the resonant wavelength of a high Q-factor mode ($< 10^3$) resulting from a quasi-bound state in the continuum formed in a LN high-contrast grating. While the amplitude variation is quite small, the variation in the phase response at a constant wavelength is used in an interferometric experiment that demonstrates the dynamic behavior of the structure at a modulation frequency of 1 MHz and a peak-to-peak voltage of 300 V. A different approach was followed in Ref. [36], where an unstructured thin-film of LN supports silicon-based nanoresonators. The variation of the refractive index of the LN layer results in a modification of the field radiated over the diffraction orders of the structures. High Q-factor modes ($> 10^4$) are designed and arranged in a device to realize electrically tunable beam splitting with an applied voltage of 30 V. More recently a first realization of a monolithic two-dimensional array nanodisks of LN obtained by partial etch of a 500 nm thick LN layer was employed to demonstrate fast electro-optic modulation of the linear transmittivity [37]. In this structure, a weak refractive index modulation of the substrate (about $10^{-5}$) resulted in an amplitude modulation of 0.01% up to 2.5 MHz for peak-to-peak voltages below 1 V. The detectable but yet small modulation is associated with the limited Q-factor attained in 2D arrangements (~$10^2$) and to the applied electric field lines that are mainly restrained in the unstructured LN film with limited spatial overlap with the optical electric field of the resonance. In Ref. [129] the resonant frequency of a plasmonic metasurface on a thin LN film is modulated. At variance from previous work, the modulating RF field lines are transverse to the metasurface plane thus allowing a small distance between the contacts, that is independent on the metasurface area. By applying a peak-to-peak voltage of 24 V, the reflectance was modulated by 40% with a modulation bandwidth of almost 800 kHz.

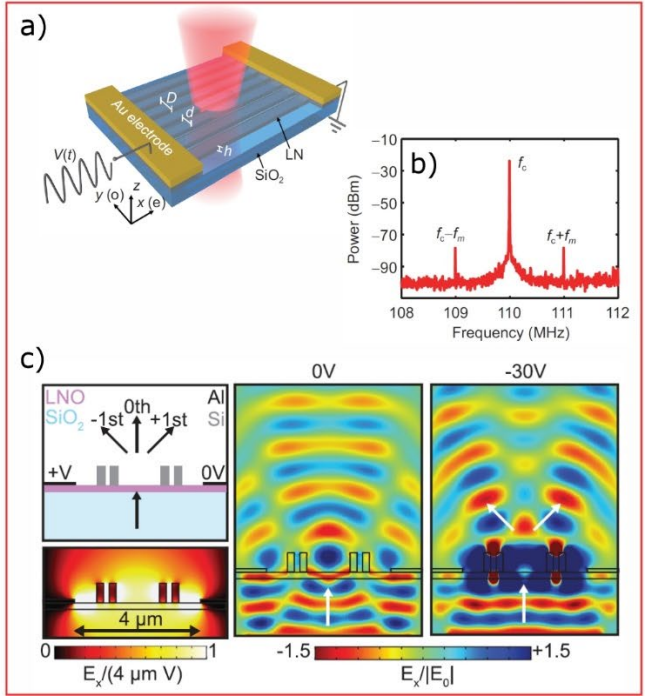
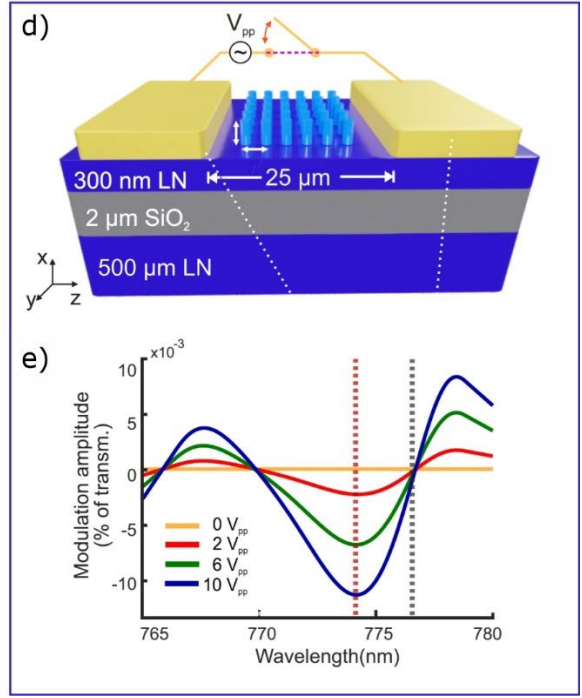

**Figure 4:** (a) LN on silica grating sustaining a quasi-bound state in the continuum between Au electrodes [35]. (b) Power spectrum of optical beats recorded with three distinct peaks are observed at $f_c - f_m$=109 MHz, $f_c$=110 MHz and $f_c + f_m$=111 MHz. (c) High Q-factor beam-splitting device [36]. Left: schematic of the device. Bottom: simulated DC electric field distribution when 1 V is applied to the left contact, keeping the right at ground. Middle: device operating at 1397 nm with no applied field. Right: device operating at 1397 nm at -30 V. (d) Metasurface (light blue LN pillars, dark blue LN bulk layer) between two Au electrodes on top of a silica layer (grey) [37]. (e) Dependence of the modulation amplitude on peak-to-peak AC voltage between 0 and 10 $V_{pp}$ ($V_{pp}$ = 1 V) with 180 kHz driving frequency.

| Reference | Q-factor | Dimension (μm²) | $V_{pp}$ (V) | Modulation rate (MHz) | Wavelength (nm) |
|---|---|---|---|---|---|
| B.F. Gao, et al., [35] | <$10^3$ | 10 × 10 | 300 | 1 | 633 |
| D. Barton, et al., [36] (only numerical) | 2.8×$10^4$ | 4 μm, not specified in the other dimension | 30 | N.A. | 1397 |
| H. Weigand, et al., [37] | 129 | 20 × 20 | 10 | 2.5 | 774.2 |
| A.Weiss, et al., [129] | 70 | 130 × 130 | 24 | 0.8 | 1535 |

**Table 4.** Summary of experimentally measured and numerically calculated electro-optical modulation by LN metasurfaces.

## 5. Perspective applications of lithium niobate meta-optics

### 5.1. Nonlinear sensing, microscopy, detectors and sources

The impressive advances reported thus far in the field of LN-based metasurfaces are rooted in the recent technological advances in nanofabrication approaches to LN. At the same time, we believe that the latest conceptual and experimental achievements in the field of meta-optics will further nourish the efforts of researchers in the field of nanofabrication, creating a positive feedback effect. We therefore expect that these first important milestones will significantly boost the development of LN-based platforms for active meta-optics spanning the whole visible range up to the mid infrared. For example, LN-based meta-optics will find a promising ground in the ever-expanding field of molecular sensing, where LN chips and waveguides have been already applied successfully for optical [130] [131] [132] and electro-optical sensing [133]. In fact, the enhanced integration enabled by LN meta-optics along with the exploitation of the ultra-narrow optical resonances (*e.g.*, quasi-BIC modes) and the nonlinear character of the generated signal employed as a probe can potentially enhance the sensitivity and scalability of these platforms.

In addition, the phase engineering of the nonlinear emission by the metasurface may allow attaining nonlinear wavefront shaping, hence steering, and even focusing *e.g.*, the SHG (see Figure 5a). This would enable the realization of nonlinear phase plates and metalenses [134]. The latter may offer a new route to nonlinear microscopy. For example nonlinear metalenses can be employed to collect visible light emitted by biological samples and down-convert it into infrared light, which can be propagated within the biological environment without experiencing significant losses. Conversely, they can be applied to upconvert infrared propagating beams into visible-UV light for in-situ microscopy.

Among the number of applications that were envisioned for nonlinear metasurfaces, one of the most fascinating and appealing is enhanced night vision. Although still based on proof of principle applications involving mainly ultrashort pulses and, hence, far from a practical use, the possibility to upconvert light from the infrared to the visible range, where cheap detectors and even the naked eye can be employed for detection, is extremely appealing. Some recent works demonstrated the possibility of translating infrared radiation to the central region of the visible spectrum by sum- frequency generation [135] [136] and third-harmonic generation [137] by means of dielectric metasurface and nanostructures. Given its broad transparency spectral window, LN-based nonlinear meta-optics offers very interesting perspectives if applied for this specific purpose (see Figure 5b).

Along with the unique properties previously mentioned, LN features lattice vibrational modes (optical phonons) in the THz range that can strongly couple to electromagnetic fields originating from phonon polaritons (PhPs) [138] [139]. While LN has been already employed for THz management as a bulk crystal [140] or in thin films [141] in the THz region (< 1 THz), the operation in the THz gap region (1–10 THz) is hindered in waveguide-based systems due to strong intrinsic absorptions. The employment of LN-based metasurfaces may ensure efficient THz generation by optical rectification in this region, which is extremely strategic to investigate PhPs in crystalline materials [142] and for the spectroscopy of macromolecules [142] [143] [144]. We also expect a distinct impact of these technologies, thanks to promising applications of THz radiation in biology and medicine [145] [146] and for the detection of explosives, weapons and drugs in security applications [147]. In this framework, very recently, a first seminal work demonstrated the capability of LN-based metasurfaces to efficiently generate THz radiation over a broadband around 1 THz [148]. Yet, the THz gap region remains widely unexplored. The technological advances in LN nanofabrication together with its high nonlinear coefficient and unique phononic features set a unique starting point for the development of THz sources seeded by optical rectification based on LN metasurfaces (see Figure 5c).

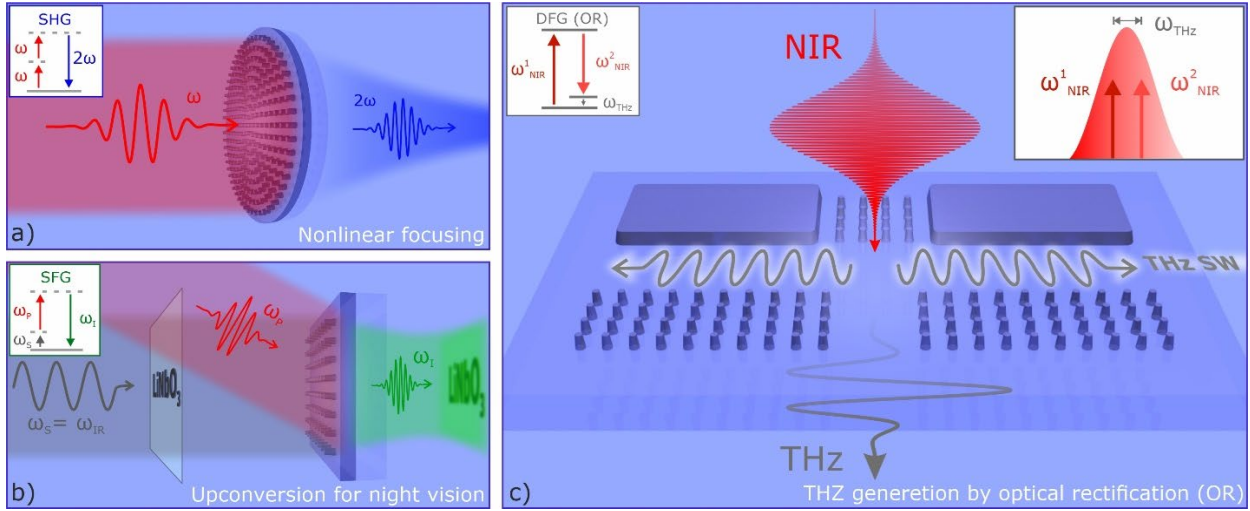

**Figure 5: perspectives and applications of nonlinear metasurfaces based on LN.** (a) Sketch of an LN monolithic metalens that focuses the nonlinear generated light (e.g., second harmonic generation, SHG) onto a sample, for noninvasive coherent imaging. (b) Sketch of a LN-based metasurface designed to upconvert infrared radiation to the visible range by SFG with a second visible light beam for night vision application [similar to [135]]. (c) Concept of a nonlinear LN metasurface for the generation of THz radiation by optical rectification (i.e., DFG), exploiting a doubly-resonant mechanism based on the Mie/lattice resonances of the meta-atom array and the design of a THz dipole meta-antenna. This concept exploits also the fact that in the region between 5 and 15 THz, LN features strong material resonances due to the presence of phonon-polaritons.

## 5.2. High-dimensional quantum-state generation

Metasurfaces are very promising to control the parameters of the transmitted light, which in quantum optics enables advanced projective measurement schemes [119] [149] [150] (see Figure 6). Particularly in LN metasurfaces, this potential can be coupled with the ability to generate photon pairs by SPDC to create complex tailored photonic quantum states.

After the first success of photon-pair generation, a lot of efforts have been dedicated to demonstrate entanglement. Recently, polarization entanglement in subwavelength films was shown [151]. Metasurface-based sources are a promising system to create entangled or even hyperentangled states, where different degrees of freedom, such as orbital angular momentum, propagation direction, wavelength, and polarization, can be simultaneously entangled thanks to the high flexibility of control granted by meta-optics. Furthermore, as every single meta-atom in a metasurface can be considered as an individual source of photon pairs independently addressable, very high-dimensional photon-pair states become possible. The potential of this approach has been demonstrated already by combining a metasurface to control the pump beam with a bulk nonlinear crystal [116]. Such high-dimensional states are an important resource for quantum information processing applications. However, to unlock these possibilities, a number of conceptional and experimental challenges has to be solved.

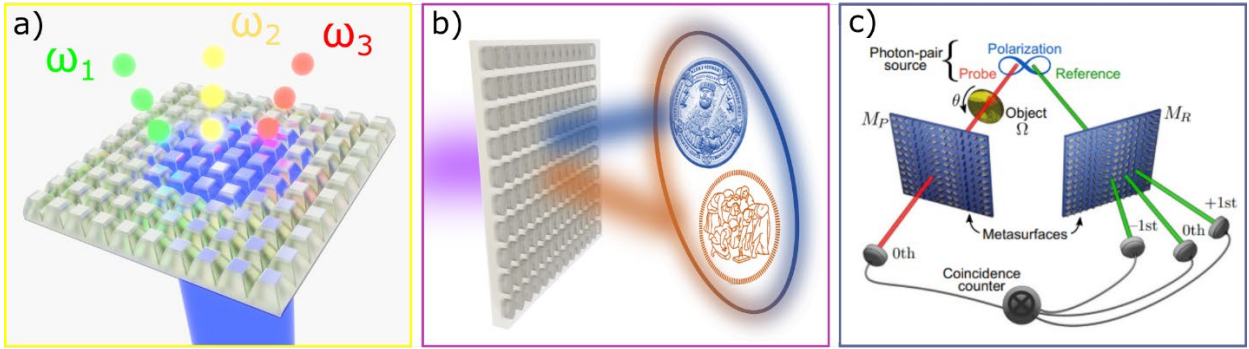

**Figure 6: perspectives and applications of metasurfaces based on LN for quantum optics** (a) Metasurface for single photon generation. b) Metasurface creating images with entangled photons. c) Metasurface for polarization ghost imaging [149] [150]. Panel c is taken from [149].

In particular, for a dedicated design of multi-parameter metasurfaces for quantum-state control, appropriate methods to calculate quantum processes in nanostructures both analytically and numerically are required. The methods typically used in integrated or bulk optics can only be partially applied to metasurfaces, which are low-Q open resonator systems with absorption. To date, among the various tools that are being developed, quantum-classical correspondence between SPDC and SFG was used to assess first experimental results [108] [152] [121] [122] and suggest new designs [117] [120]. A more rigorous approach is based on the Green's function method [153] for constructing two-photon amplitudes, which was shown to be able to calculate efficiency and directionality of photon-pair generation from a single spherical nanoresonator [154]. But this method is hardly applicable for metasurfaces as it requires a rigorous description of the modal composition in terms of multipoles. A more general approach also based on Green's functions uses the quasi-normal modes formalism to describe nonlinear nanoresonators [95] and has great potential to describe SPDC in metasurfaces [155].

Yet, the current rate of generated photon pairs by SPDC in metasurfaces is far from being of any practical use because of the low efficiency of the SPDC process. Clever designs based on BIC resonances or coupled Mie/BIC modes with Bragg resonances can boost the electromagnetic fields inside the resonators by improving the coupling with the resonators and/or increase effective interaction length [156]. Another strategy consists in exploiting hybrid structures, where a metasurface from another material couples electromagnetic field into an LN slab, as recently demonstrated by Zhang et al. [108]. Controlling the SPDC emission direction is yet another way to increase detected photon pairs, where optimized design can help to improve the directionality so that more photons can reach the detectors. To date, particular efforts towards this direction are still required.

Besides photon pairs, also single photons are often used as non-classical states of light, which can profit from the ability of metasurfaces for complex manipulations. Single photons are typically generated with atom-like emitters [157] [158] [159]. Although pure LN does not possess the color centers typically employed for this task in solid state systems, selective doping with rare-earth ions (Er, Tm, Yb) could provide such emitters. Moreover, LN can be integrated with III-V semiconductor quantum dots [73] or transition metal dichalcogenides [160] emitting single photons. Here, LN is a very suitable material that, due to its large transparency range, is compatible with many different emitters. Furthermore, one can take advantage of heralding to obtain single photons from photon pairs generated by SPDC, where one photon from a pair is used for quantum applications, while the second photon "heralds" its presence.

We anticipate that in the coming years quantum optics on the nanoscale will rapidly develop and give birth to many new devices and applications. Of course, other materials are considered for such tasks as well [161], but LN is one of the most promising platforms due to its combination of properties.

### 5.3. Future electro-optical modulators

The use of electro-optic effects to achieve dynamic control the optical response from LN nanostructures and metasurfaces has just began. From these first works it is clear that the main challenge resides into enhancing the electric field to sensibly exploit the relatively small refractive index change induced by the Pockels effect. Therefore, future developing trends might focus on engineering the spatial profiles of the high-Q optical resonant modes to increase the overlap with applied electric field lines inside the metasurface volume. From a fabrication perspective, the capability to locate the electrical contacts on top and bottom of sub-micrometer thin resonators will likely benefit the modulation strength and reduce the operating voltage. Harnessing the electro-optic effect in LN nanostructures and metasurfaces may pave the way for the development of a new generation of ultrafast integrated devices such as tunable optical lenses and high-resolution spatial light modulators.

## 6. Conclusions

We have reviewed the most recent advances in the field of LN meta-optics, discussing potential perspectives for applications in active light management and quantum optics, with a critical eye on the potential setbacks and limitations of the LN platform with respect to materials and platforms devised in the emerging field of all-dielectric active meta-optics. Recent advances in nanofabrication technology indeed allowed the miniaturization of LN platforms, hence enabling the demonstration of LN-based metasurfaces. Thanks to the unique properties of LN it is now possible to envision the realization of ultra-flat monolithic nonlinear light sources with emission ranging from the visible to the infrared, efficient sources of correlated photon pairs as well as electro-optical devices. Despite the undoubtable advantages of the material, its relatively low refractive index limits the field confinement compared to other dielectric platforms, hence potentially hindering the nonlinear optic and electro-optical performance of LN-based active meta-optics. Yet, the successful recent realization of nonlinear light conversion, photon pair generation and electro-optical modulation via LN metasurfaces will serve as trigger to further improve of LN fabrication, hence allowing the realization of high-Q metasurfaces, which might be the key landmark to take LN-based active meta-optics towards relevant technological applications.

3232